\documentclass[showpacs,preprintnumbers,amsmath,amssymb]{revtex4}

\usepackage{graphicx}
\usepackage{dcolumn}
\usepackage{bm}

\newcommand{\bq}{\begin{equation}}
\newcommand{\eq}{\end{equation}}
\newcommand{\bqa}{\begin{eqnarray}}
\newcommand{\eqa}{\end{eqnarray}}
\newcommand{\nn}{\nonumber \\}

\def\be     {\begin{equation}}
\def\ee     {\end{equation}}
\def\bea        {\begin{eqnarray}}
\def\eea        {\end{eqnarray}}
\def\bnn    {\begin{eqnarray*}}
\def\enn    {\end{eqnarray*}}

\begin{document}

\title{Beyond quantum chaos in emergent dual holography}

\author{Ki-Seok Kim$^{a,b}$}
\affiliation{$^{a}$Department of Physics, POSTECH, Pohang, Gyeongbuk 37673, Korea \\ $^{b}$Asia Pacific Center for Theoretical Physics (APCTP), Pohang, Gyeongbuk 37673, Korea}

%
%

%
\email[Ki-Seok Kim: ]{tkfkd@postech.ac.kr}
%
%

\date{\today}

\begin{abstract}
Black hole is well known to be a fast scrambler, responsible for physics of quantum chaos in dual holography. Recently, the Euclidean worm hole has been proposed to play a central role in the chaotic behavior of the spectral form factor. Furthermore, this phenomena was reinterpreted based on an effective field theory approach for quantum chaos. Since the graded nonlinear $\sigma-$model approach can describe not only the Wigner-Dyson level statistics but also its Poisson distribution, it is natural to ask whether the dual holography can touch the Poisson regime beyond the quantum chaos. In this study, we investigate disordered strongly coupled conformal field theories in the large central-charge limit. An idea is to consider a quenched average for metric fluctuations and to take into account the renormalization group flow of the metric-tensor distribution function from the UV to the IR boundary. Here, renormalization effects at a given disorder configuration are described by the conventional dual holography. We uncover that the renormalized distribution function shows a power-law behavior universally, interpreted as an infinite randomness fixed point.
\end{abstract}

\maketitle

\section{Introduction}

Quantum chaos \cite{Quantum_Chaos_Review} and effective hydrodynamics \cite{Holographic_Liquid_Son_I,Holographic_Liquid_Son_II,Holographic_Liquid_Son_III,Holographic_Liquid_Son_IV} are two fingerprints in dual holography \cite{Holographic_Duality_I,Holographic_Duality_II,Holographic_Duality_III,Holographic_Duality_IV}. Black hole \cite{GR_Textbook} plays a central role in this physics, regarded to be a fast scrambler \cite{Black_Hole_Fast_Scrambler} and reflected in the entropy formula proportional to its area \cite{Black_Hole_Entropy_I,Black_Hole_Entropy_II}. Such extensive degrees of freedom serve as the source of strong inelastic scattering, responsible for the quantum chaos and the emergent hydrodynamics.

Recently, the quantum chaos in the dual holography has been more deeply understood by the study of the spectral form factor \cite{Black_Hole_Random_Matrix}. It turns out that the Euclidean worm hole geometry was proposed to be responsible for the quantum chaos in the spectral form factor perspective, more precisely, the $t-$linear increase (`ramp' behavior) in the spectral form factor \cite{Euclidean_Wormhole}. Through the Euclidean worm hole, the spectra of both boundary conformal field theories are correlated, giving rise to so called the phenomena of level repulsion \cite{Sonner_QC_EFT_I}.

The level-repulsion phenomena has been revisited in an effective field theory approach. The spectral form factor is reformulated in a graded nonlinear $\sigma-$model field theory, reproducing the ramp behavior in a perturbative approach of the large Hilbert-space dimensional limit \cite{Sonner_QC_EFT_II}. In particular, this graded nonlinear $\sigma-$model approach tries to construct correspondences in order by order between the field theory approach and the quantum gravity theory of the dual holography, where the role of the Euclidean worm hole is more clarified.

We would like to point out that the supersymmetric $\sigma-$model approach \cite{SUSY_NLsM_Review} can describe not only the Wigner-Dyson level statistics but also its Poisson distribution, which allows level degeneracy. If this effective field theory is dual to the gravity theory of the holographic approach, it is natural to ask whether the dual holography can touch the Poisson-distribution regime beyond the quantum chaos identified with the Wigner-Dyson distribution. We believe that this is a physically reasonable question that we address in this paper.

In this study, we investigate disordered strongly coupled conformal field theories in the large central-charge limit. We recall that the Wigner-Dyson distribution of level statistics appears in a diffusive metallic phase while the Poisson level statistics arises in an Anderson insulating state \cite{Anderson_Localization_Review}. In this respect we try to reach an analogue of the Anderson localized phase in the dual holography. An idea is to consider a quenched average for metric fluctuations and to take into account the renormalization group flow of the metric-tensor distribution function from the UV to the IR boundary \cite{Strong_Disorder_RG}. Here, renormalization effects at a given disorder configuration are described by the conventional dual holography \cite{Holographic_Duality_V,Holographic_Duality_VI,Holographic_Duality_VII}. It turns out that the renormalized distribution function shows a power-law behavior universally. We interpret that this power-law distribution function describes an infinite randomness fixed point \cite{Strong_Disorder_RG} in strongly coupled disordered conformal field theories of the large central-charge limit.

\section{General prescription for strongly disordered and interacting electrons}

Before going to a concrete example, we discuss a general structure of this problem. We start from the following free energy
\bqa && \mathcal{F} = - \frac{1}{\beta} \int_{-\infty}^{\infty} d v_{b}(x) P[\{v_{b}(x)\}] \ln \int D c_{\sigma}(x,\tau) \exp\Big\{ - \int_{0}^{\beta} d \tau \int d^{d} x \mathcal{L}[c_{\sigma}(x,\tau);\{g_{a}(x,\tau)\},\{v_{b}(x)\}] \Big\} . \eqa
Here, $\mathcal{L}[c_{\sigma}(x,\tau);\{g_{a}(x,\tau)\},\{v_{a}(x)\}]$ is an effective UV Lagrangian to describe dynamics of electrons $c_{\sigma}(x,\tau)$ under their effective interactions with strengths $\{g_{a}(x,\tau)\}$ and quenched random potentials with strengths $\{v_{a}(x)\}$. $\sigma$ represents the electron spin, extended from $2$ to $N$. $a$ and $b$ denote the number of effective interactions and random potentials, respectively.
\bqa && F[\{v_{b}(x)\}] = - \frac{1}{\beta} \ln \int D c_{\sigma}(x,\tau) \exp\Big\{ - \int_{0}^{\beta} d \tau \int d^{d} x \mathcal{L}[c_{\sigma}(x,\tau);\{g_{a}(x,\tau)\},\{v_{b}(x)\}] \Big\} \nonumber \eqa
is the free energy, given a disorder configuration $\{v_{b}(x)\}$ at UV. $\beta$ is inverse temperature. To obtain the physically observable free energy, we have to perform quenched-averaging of the free energy for various realizations of disorder configurations. Here, $P[\{v_{b}(x)\}]$ is the probability distribution function of disorder configurations, and
\bqa && \mathcal{F} = \int_{-\infty}^{\infty} d v_{b}(x) P[\{v_{b}(x)\}] F[\{v_{b}(x)\}] \nonumber \eqa
is the quenched-averaged effective free energy. $\int_{-\infty}^{\infty} d v_{b}(x) P[\{v_{b}(x)\}] = 1$ is assumed.

Now, we perform a functional renormalization group transformation in real space. Although it is not easy to take such a task, we perform the Kadanoff block-spin transformation explicitly in one dimension \cite{Kadanoff_RG}, to be presented below. In higher dimensions, it is more difficult to perform the functional renormalization group transformation because the lattice structure is modified to depend on the renormalization-group transformation step.
%
%
See refs. \cite{SungSik_Holography_I,SungSik_Holography_II,SungSik_Holography_III, Einstein_Klein_Gordon_RG_Kim,Einstein_Dirac_RG_Kim,RG_GR_Geometry_I_Kim,RG_GR_Geometry_II_Kim,RG_GR_Geometry_III_Kim} for real-space renormalization group transformations above one spatial dimension. Whatever the regularization is, suppose that we take the renormalization group transformation. Then, we would obtain the following free energy
\bqa && \mathcal{F} = - \frac{1}{\beta} \int_{-\infty}^{\infty} d g_{a}(x,\tau,z_{f}) d v_{a}(x,z_{f}) P[\{g_{a}(x,\tau,z_{f})\}, \{v_{a}(x,z_{f})\};z_{f}] \nn && \ln \int D c_{\sigma}(x,\tau) D g_{a}(x,\tau,z) D v_{a}(x,z) \delta\Big(v_{a}(x,0) - v_{a}(x)\Big) \delta\Big(g_{a}(x,\tau,0) - g_{a}(x,\tau)\Big) \nn && \delta\Big\{ \partial_{z} v_{a}(x,z) - \beta_{v_{a}}[\{g_{a}(x,\tau,z)\}, \{v_{a}(x,z)\}] \Big\} \mbox{det} \Big\{ \partial_{z} \delta_{ab} - \frac{\partial }{\partial v_{b}(x,z)} \beta_{v_{a}}[\{g_{a}(x,\tau,z)\}, \{v_{a}(x,z)\}] \Big\} \nn && \delta\Big\{ \partial_{z} g_{a}(x,\tau,z) - \beta_{g_{a}}[\{g_{a}(x,\tau,z)\}, \{v_{a}(x,z)\}] \Big\} \mbox{det} \Big\{ \partial_{z} \delta_{ab} - \frac{\partial }{\partial g_{b}(x,z)} \beta_{g_{a}}[\{g_{a}(x,\tau,z)\}, \{v_{a}(x,z)\}] \Big\} \nn && \exp\Big\{ - \int_{0}^{\beta} d \tau \int d^{d} x \mathcal{L}[c_{\sigma}(x,\tau);\{g_{a}(x,\tau,z_{f})\},\{v_{a}(x,z_{f})\}] - N \int_{0}^{z_{f}} d z \int_{0}^{\beta} d \tau \int d^{d} x \mathcal{V}_{eff}[\{g_{a}(x,\tau,z)\},\{v_{a}(x,z)\}] \Big\} . \nn \eqa
Here,
\bqa && F[\{g_{a}(x,\tau,z_{f})\}, \{v_{a}(x,z_{f})\}] = - \frac{1}{\beta} \ln \int D c_{\sigma}(x,\tau) D g_{a}(x,\tau,z) D v_{a}(x,z) \delta\Big(v_{a}(x,0) - v_{a}(x)\Big) \delta\Big(g_{a}(x,\tau,0) - g_{a}(x,\tau)\Big) \nn && \delta\Big\{ \partial_{z} v_{a}(x,z) - \beta_{v_{a}}[\{g_{a}(x,\tau,z)\}, \{v_{a}(x,z)\}] \Big\} \mbox{Det} \Big\{ \partial_{z} \delta_{ab} - \frac{\partial }{\partial v_{b}(x,z)} \beta_{v_{a}}[\{g_{a}(x,\tau,z)\}, \{v_{a}(x,z)\}] \Big\} \nn && \delta\Big\{ \partial_{z} g_{a}(x,\tau,z) - \beta_{g_{a}}[\{g_{a}(x,\tau,z)\}, \{v_{a}(x,z)\}] \Big\} \mbox{Det} \Big\{ \partial_{z} \delta_{ab} - \frac{\partial }{\partial g_{b}(x,z)} \beta_{g_{a}}[\{g_{a}(x,\tau,z)\}, \{v_{a}(x,z)\}] \Big\} \nn && \exp\Big\{ - \int_{0}^{\beta} d \tau \int d^{d} x \mathcal{L}[c_{\sigma}(x,\tau);\{g_{a}(x,\tau,z_{f})\},\{v_{a}(x,z_{f})\}] - N \int_{0}^{z_{f}} d z \int_{0}^{\beta} d \tau \int d^{d} x \mathcal{V}_{eff}[\{g_{a}(x,\tau,z)\},\{v_{a}(x,z)\}] \Big\} \nn \eqa
is an effective free energy functional in terms of both effective interactions $\{g_{a}(x,\tau,z_{f})\}$ and random potentials $\{v_{a}(x,z_{f})\}$ renormalized at IR, given a disorder configuration $\{v_{b}(x)\}$ at UV. 
\bqa && N \int_{0}^{z_{f}} d z \int_{0}^{\beta} d \tau \int d^{d} x \mathcal{V}_{eff}[\{g_{a}(x,\tau,z)\},\{v_{a}(x,z)\}] \nn && = - \frac{1}{\beta} \ln \int_{\Lambda(z)} D c_{\sigma}(x,\tau) \exp\Big\{ - \int_{0}^{\beta} d \tau \int d^{d} x \mathcal{L}[c_{\sigma}(x,\tau);\{g_{a}(x,\tau,z)\},\{v_{a}(x,z)\}] \Big\} \eqa
is an effective potential at a given energy scale $z$ with a reduced cutoff $\Lambda(z)$ for $c_{\sigma}(x,\tau)$, which arises from the renormalization group transformation at the scale of $z$. $z$ is the renormalization-group transformation scale. 

Both disorders $\{v_{a}(x,z)\}$ and interactions $\{g_{a}(x,\tau,z)\}$ evolve from their UV bare values of $v_{a}(x,0) = v_{a}(x)$ and $g_{a}(x,\tau,0) = g_{a}(x,\tau)$ to their IR renormalized ones of $\{v_{a}(x,z_{f})\}$ and $\{g_{a}(x,\tau,z_{f})\}$, respectively, through the renormalization group transformation, given by their renormalization group flows
\bqa && \partial_{z} v_{a}(x,z) = \beta_{v_{a}}[\{g_{a}(x,\tau,z)\}, \{v_{a}(x,z)\}] , ~~~~~ \partial_{z} g_{a}(x,\tau,z) = \beta_{g_{a}}[\{g_{a}(x,\tau,z)\}, \{v_{a}(x,z)\}] . \eqa
Here, $\beta_{v_{a}}[\{g_{a}(x,\tau,z)\}, \{v_{a}(x,z)\}]$ and $\beta_{g_{a}}[\{g_{a}(x,\tau,z)\}, \{v_{a}(x,z)\}]$ are renormalization group $\beta-$functions \cite{QFT_Textbook}, resulting from high-energy quantum fluctuations of matter fields. More precisely, they are given by the effective potential as follows
\bqa && \beta_{v_{a}}[\{g_{a}(x,\tau,z)\}, \{v_{a}(x,z)\}] = - \frac{\partial}{\partial v_{a}(x,z)} \mathcal{V}_{eff}[\{g_{a}(x,\tau,z)\},\{v_{a}(x,z)\}] , \\ && \beta_{g_{a}}[\{g_{a}(x,\tau,z)\}, \{v_{a}(x,z)\}] = - \frac{\partial}{\partial g_{a}(x,z)} \mathcal{V}_{eff}[\{g_{a}(x,\tau,z)\},\{v_{a}(x,z)\}] . \eqa
Such fully renormalized coupling functions and random potentials appear in the IR renormalized effective Lagrangian $\mathcal{L}[c_{\sigma}(x,\tau);\{g_{a}(x,\tau,z_{f})\},\{v_{a}(x,z_{f})\}]$ for a given disorder realization at UV. $\mbox{Det} \Big\{ \partial_{z} \delta_{ab} - \frac{\partial }{\partial v_{b}(x,z)} \beta_{v_{a}}[\{g_{a}(x,\tau,z)\}, \{v_{a}(x,z)\}] \Big\}$ and $\mbox{Det} \Big\{ \partial_{z} \delta_{ab} - \frac{\partial }{\partial g_{b}(x,z)} \beta_{g_{a}}[\{g_{a}(x,\tau,z)\}, \{v_{a}(x,z)\}] \Big\}$ are Jacobian factors, which appear in the Faddeev-Popov procedure \cite{QFT_Textbook}. In other words, we obtain
\bqa && \int D g_{a}(x,\tau,z) D v_{a}(x,z) \delta\Big\{ \partial_{z} v_{a}(x,z) - \beta_{v_{a}}[\{g_{a}(x,\tau,z)\}, \{v_{a}(x,z)\}] \Big\} \mbox{Det} \Big\{ \partial_{z} \delta_{ab} - \frac{\partial }{\partial v_{b}(x,z)} \beta_{v_{a}}[\{g_{a}(x,\tau,z)\}, \{v_{a}(x,z)\}] \Big\} \nn && \delta\Big\{ \partial_{z} g_{a}(x,\tau,z) - \beta_{g_{a}}[\{g_{a}(x,\tau,z)\}, \{v_{a}(x,z)\}] \Big\} \mbox{Det} \Big\{ \partial_{z} \delta_{ab} - \frac{\partial }{\partial g_{b}(x,z)} \beta_{g_{a}}[\{g_{a}(x,\tau,z)\}, \{v_{a}(x,z)\}] \Big\} = 1 , \eqa
which leads the partition function to be invariant under the renormalization group transformation. 

Introducing Lagrange multiplier and ghost fields into the above partition function, we reformulate the effective field theory as follows
\bqa && \mathcal{F} = - \frac{1}{\beta} \int_{-\infty}^{\infty} d g_{a}(x,\tau,z_{f}) d v_{a}(x,z_{f}) P[\{g_{a}(x,\tau,z_{f})\}, \{v_{a}(x,z_{f})\};z_{f}] \nn && \ln \int D c_{\sigma}(x,\tau) D g_{a}(x,\tau,z) D \Pi_{g_{a}}(x,z) D v_{a}(x,z) D \Pi_{v_{a}}(x,z) D \bar{c}_{a}(x,z) D c_{a}(x,z) D \bar{f}_{a}(x,z) D f_{a}(x,z) \nn && \delta\Big(v_{a}(x,0) - v_{a}(x)\Big) \delta\Big(g_{a}(x,\tau,0) - g_{a}(x,\tau)\Big) \exp\Big[ - \int_{0}^{\beta} d \tau \int d^{d} x \mathcal{L}[c_{\sigma}(x,\tau);\{g_{a}(x,\tau,z_{f})\},\{v_{a}(x,z_{f})\}] \nn && - N \int_{0}^{z_{f}} d z \int_{0}^{\beta} d \tau \int d^{d} x \Big\{ \Pi_{v_{a}}(x,z) \Big(\partial_{z} v_{a}(x,z) - \beta_{v_{a}}[\{g_{a}(x,\tau,z)\}, \{v_{a}(x,z)\}]\Big) \nn && + \Pi_{g_{a}}(x,z) \Big(\partial_{z} g_{a}(x,\tau,z) - \beta_{g_{a}}[\{g_{a}(x,\tau,z)\}, \{v_{a}(x,z)\}]\Big) + \mathcal{V}_{eff}[\{g_{a}(x,\tau,z)\},\{v_{a}(x,z)\}] \nn && + \bar{c}_{a}(x,z) \Big(\partial_{z} \delta_{ab} - \frac{\partial }{\partial v_{b}(x,z)} \beta_{v_{a}}[\{g_{a}(x,\tau,z)\}, \{v_{a}(x,z)\}]\Big) c_{b}(x,z) \nn && + \bar{f}_{a}(x,z) \Big(\partial_{z} \delta_{ab} - \frac{\partial }{\partial g_{b}(x,z)} \beta_{g_{a}}[\{g_{a}(x,\tau,z)\}, \{v_{a}(x,z)\}]\Big) f_{b}(x,z) \Big\} \Big] . \eqa
This expression manifests the renormalization group transformation in the level of an effective action, claimed to be an emergent holographic dual effective field theory. Although it would be interesting to discuss the formal aspect of this effective field theory more deeply, we do not discuss symmetries and Ward identities further and focus on the disorder physics in this paper.

An idea is to introduce the renormalization group flow for the distribution function of random potentials and effective interactions into the quenched average of the free energy as follows
\bqa && \mathcal{F} = \int_{-\infty}^{\infty} d g_{a}(x,\tau,z_{f}) d v_{a}(x,z_{f}) P[\{g_{a}(x,\tau,z_{f})\}, \{v_{a}(x,z_{f})\};z_{f}] F[\{g_{a}(x,\tau,z_{f})\}, \{v_{a}(x,z_{f})\}] . \eqa
Here, $P[\{g_{a}(x,\tau,z_{f})\}, \{v_{a}(x,z_{f})\};z_{f}]$ is the renormalized distribution function for renormalized interactions and potentials at IR. Considering that the free energy functional $F[\{g_{a}(x,\tau,z_{f})\}, \{v_{a}(x,z_{f})\}]$ has to be invariant under the renormalization group transformation in a given disorder configuration at UV, we obtain the following identity
\bqa && \int_{-\infty}^{\infty} d g_{a}(x,\tau,z_{f}) d v_{a}(x,z_{f}) P[\{g_{a}(x,\tau,z_{f})\}, \{v_{a}(x,z_{f})\};z_{f}] \nn && = \int_{-\infty}^{\infty} d g_{a}(x,\tau,z_{f}+dz) d v_{a}(x,z_{f}+dz) P[\{g_{a}(x,\tau,z_{f}+dz)\}, \{v_{a}(x,z_{f}+dz)\};z_{f}+dz] . \eqa
As a result, we obtain
\bqa && P[\{ \lambda_{a}(x,\tau,z) \};z] = \mbox{Det} \begin{pmatrix} \delta_{ab} + d z \frac{\partial \beta_{\lambda_{a}}[\{ \lambda_{a}(x,\tau,z) \}]}{\partial \lambda_{b}(x,\tau,z)} \end{pmatrix} P[\{ \lambda_{a}(x,\tau,z) + d z \beta_{\lambda_{a}}[\{ \lambda_{a}(x,\tau,z)\}] \};z + d z] , \label{RG_Flow_Mother} \eqa
where the IR energy scale $z_{f}$ is replaced with a general scale $z$. Here, we used a short-hand notation for $\lambda_{a}(x,\tau,z) \equiv (g_{a}(x,\tau,z), v_{a}(x,z))$ as follows
\bqa && \partial_{z} \lambda_{a}(x,\tau,z) = \partial_{z} (g_{a}(x,\tau,z), v_{a}(x,z)) \nn && = (\beta_{g_{a}}[\{g_{a}(x,\tau,z)\}, \{v_{a}(x,z)\}], \beta_{v_{a}}[\{g_{a}(x,\tau,z)\}, \{v_{a}(x,z)\}]) \equiv \beta_{\lambda_{a}}[\{ \lambda_{a}(x,\tau,z) \}] . \nonumber \eqa

Physical meaning of Eq. (\ref{RG_Flow_Mother}) is simple. We just rewrite the distribution function of old variables $\{ \lambda_{a}(x,\tau,z) \}$ at the energy scale $z$ as that of renormalization-group transformation updated ones $\{ \lambda_{a}(x,\tau,z) + \beta_{\lambda_{a}}[\{ \lambda_{a}(x,\tau,z)\}] \}$ at the energy scale $z + d z$. Here, $\beta_{\lambda_{a}}[\{ \lambda_{a}(x,\tau,z)\}]$ is the renormalization-group $\beta-$function. $\mbox{Det} \begin{pmatrix} \delta_{ab} + \frac{\partial \beta_{\lambda_{a}}[\{ \lambda_{a}(x,\tau,z) \}]}{\partial \lambda_{b}(x,\tau,z)} \end{pmatrix}$ is nothing but the Jacobian factor to count the change of the `volume' integration.

It is straightforward to reformulate Eq. (\ref{RG_Flow_Mother}) as the following differential equation
\bqa && \Big\{ \frac{\partial}{\partial z} + \beta_{\lambda_{a}}[\{ \lambda_{a}(x,\tau,z) \}] \frac{\partial}{\partial \lambda_{a}(x,\tau,z)} + \mbox{tr} \Big( \frac{\partial \beta_{\lambda_{a}}[\{ \lambda_{a}(x,\tau,z) \}]}{\partial \lambda_{b}(x,\tau,z)} \Big) \Big\} P[\{ \lambda_{a}(x,\tau,z) \}; z] = 0 , \label{Callan__Symanzik_Eq_Distribution} \eqa
which may be regarded as one of the main results in this study. This first-order differential equation can be considered as the Callan-Symanzik equation \cite{QFT_Textbook} for the distribution function of all the coupling functions although there appears the Jacobian factor due to the change of the volume integration in the coupling-function space.

It is not possible to solve Eq. (\ref{Callan__Symanzik_Eq_Distribution}) in a general situation. Instead, it is not difficult to examine this renormalization group flow near a fixed point, given by $\beta_{\lambda_{a}}[\{ \lambda_{a}^{*}(x,\tau) \}] = 0$. The linearized renormalization group $\beta-$function is given by
\bqa && \partial_{z} \delta \lambda_{a}(x,\tau,z) = \frac{\partial \beta_{\lambda_{a}}[\{ \lambda_{a}(x,\tau,z) \}]}{\partial \lambda_{b}(x,\tau,z)} \Big|_{\beta_{\lambda_{a}}[\{ \lambda_{a}^{*}(x,\tau) \}] = 0} \delta \lambda_{b}(x,\tau,z) \equiv \nu_{ab}(x,\tau) \delta \lambda_{b}(x,\tau,z) \label{Linear_Beta_Ft} \eqa
near the fixed point. Here, $\nu_{ab}(x,\tau) = \frac{\partial \beta_{\lambda_{a}}[\{ \lambda_{a}(x,\tau,z) \}]}{\partial \lambda_{b}(x,\tau,z)} \Big|_{\beta_{\lambda_{a}}[\{ \lambda_{a}^{*}(x,\tau) \}] = 0}$ is the local critical exponent.
%
%
As a result, the renormalization group flow of the distribution function reads
\bqa && \Big(\frac{\partial}{\partial z} + \nu_{ab}(x,\tau) \delta \lambda_{b}(x,\tau,z) \frac{\partial}{\partial \delta \lambda_{a}(x,\tau,z)} \Big) \ln P_{*}[\{ \delta \lambda_{a}(x,\tau,z) \}; z] = - \nu_{aa}(x,\tau) \label{Linear_RG_Flow_Distribution} \eqa
near the fixed point $\{ \lambda_{a}^{*}(x,\tau) \}$. Here, the right hand side represents trace of the critical-exponent matrix.

To solve Eq. (\ref{Linear_RG_Flow_Distribution}) with Eq. (\ref{Linear_Beta_Ft}), we diagonalize Eq. (\ref{Linear_Beta_Ft}) as follows
\bqa && \partial_{z} \delta \overline{\lambda}_{a}(x,\tau,z) = \overline{\nu}_{aa}(x,\tau) \delta \overline{\lambda}_{a}(x,\tau,z) . \eqa
$\delta \overline{\lambda}_{a}(x,\tau,z)$ forms the diagonalized basis near the fixed point, and $\overline{\nu}_{aa}(x,\tau)$ is the actual critical exponent. Then, the Callan-Symanzik equation for the distribution function is given by
\bqa && \Big(\frac{\partial}{\partial z} + \overline{\nu}_{aa}(x,\tau) \frac{\partial}{\partial \ln \delta \overline{\lambda}_{a}(x,\tau,z)} \Big) \ln P_{*}[\{ \delta \overline{\lambda}_{a}(x,\tau,z) \}; z] = - \overline{\nu}_{aa}(x,\tau) \eqa
near the fixed point. As a result, we find
%
%
\bqa && P_{*}[\{ \delta \overline{\lambda}_{a}(x,\tau,z) \}; z] = \mathcal{C}^{*} \Pi_{a = 1} \Big( [\delta \overline{\lambda}_{a}(x,\tau,z)]^{\frac{\sum_{b} \overline{\nu}_{bb}(x,\tau)}{\overline{\nu}_{aa}(x,\tau)}} \Big) , \eqa
which shows a power-law behavior, characterized by the critical exponent of the fixed point with a given disorder configuration at UV. $\mathcal{C}^{*}$ is a normalization constant.
%
%
Since this power-law distribution function gives rise to infinite variances for variables, this distribution function governs an infinite randomness fixed point \cite{Strong_Disorder_RG}.

%
%

This theoretical framework reminds us of the strong-disorder renormalization-group approach \cite{Strong_Disorder_RG}, mainly applied to one-dimensional strongly disordered systems. It is fair to say that our theoretical framework benchmarks this strong-disorder renormalization group approach actually. However, we emphasize that the renormalization group transformation in this study is taken int account in a homogeneous way instead of picking up strong disorder positions in the previous approach. Moreover, we suggest how this framework can be generalized to the dual holography, which allows us to consider higher dimensional strongly coupled systems.

\section{A functional renormalization group transformation method for one-dimensional disordered noninteracting fermions}

\subsection{Kadanoff block-spin transformation and renormalization group transformation for the distribution function of disorder}

As a proof of working principle, we consider one-dimensional disordered noninteracting fermions. The averaged free energy is given by
\bqa && \mathcal{F} = - \frac{1}{\beta} \int_{-\infty}^{\infty} d v_{i} P(v_{i}) \ln \int D c_{i \sigma} \exp\Big[ - \int_{0}^{\beta} d \tau \sum_{i = 1}^{N} \Big\{ c_{i\sigma}^{\dagger} (\partial_{\tau} - \mu + v_{i}) c_{i\sigma} - t_{i} (c_{i\sigma}^{\dagger} c_{i+1\sigma} + c_{i+1\sigma}^{\dagger} c_{i\sigma}) \Big\} \Big] . \eqa
Here, $c_{i \sigma}$ is an electron field with spin $\sigma = 1, ..., N_{s}$ on a lattice site $i$. $\mu$ and $t_{i}$ are the chemical potential and the hopping integral, respectively. Although the hopping integral defined on a link is better to be written as $t_{i i+1}$, we use a short-hand notation $t_{i}$. $v_{i}$ is a quenched random potential, the distribution of which is
\bqa && P(v_{i}) = \Big( 2 \pi \Gamma_{v} \Big)^{- \frac{N}{2}} \exp\Big\{ - \frac{1}{2 \Gamma_{v}} \sum_{i = 1}^{N} v_{i}^{2} \Big\} . \eqa
This gaussian distribution function is normalized as $\int_{-\infty}^{\infty} d v_{i} P(v_{i}) = 1$.

To perform the Kadanoff block-spin transformation in a recursive way, we introduce a superscript $(0)$, which represents the iteration number of renormalization group transformations as follows
\bqa && \mathcal{F} = - \frac{1}{\beta} \int_{-\infty}^{\infty} d \mu_{i}^{(0)} d t_{i}^{(0)} P^{(0)}[\mu_{i}^{(0)},t_{i}^{(0)}] \ln \int D c_{i \sigma} D \mu_{i}^{(0)} D t_{i}^{(0)} \delta(\mu_{i}^{(0)} - \mu + v_{i}) \delta(t_{i}^{(0)} - t_{i}) \nn && \exp\Big[ - \int_{0}^{\beta} d \tau \sum_{i = 1}^{N} \Big\{ c_{i\sigma}^{\dagger} (\partial_{\tau} - \mu_{i}^{(0)}) c_{i\sigma} - t_{i}^{(0)} (c_{i\sigma}^{\dagger} c_{i+1\sigma} + c_{i+1\sigma}^{\dagger} c_{i \sigma}) \Big\} \Big] . \eqa
Here, the expression of $\int D \mu_{i}^{(0)} D t_{i}^{(0)} \delta(\mu_{i}^{(0)} - \mu + v_{i}) \delta(t_{i}^{(0)} - t_{i})$ is easily understood. The quenched average part is rewritten as
\bqa && \int_{-\infty}^{\infty} d \mu_{i}^{(0)} P(\mu_{i}^{(0)}) = \int_{-\infty}^{\infty} d \mu_{i}^{(0)} d t_{i}^{(0)} P(\mu_{i}^{(0)}) \delta(t_{i}^{(0)} - t_{i}) \equiv \int_{-\infty}^{\infty} d \mu_{i}^{(0)} d t_{i}^{(0)} P^{(0)}[\mu_{i}^{(0)},t_{i}^{(0)}] \eqa
with $P^{(0)}[\mu_{i}^{(0)},t_{i}^{(0)}] = P(\mu_{i}^{(0)}) \delta(t_{i}^{(0)} - t_{i})$ because the renormalization group transformation in a given disorder configuration leads the distribution function to depend on the hopping integral, as will be seen below.

Separating all dynamical fields into even-site and odd-site degrees of freedom and performing the path integral with respect to even-site degrees of freedom, we obtain an effective lattice field theory on odd-site quantum fields. Then, we rescale the lattice structure coming back to the original one and all quantum fields recovering the original expression of the lattice field theory. As a result, we obtain
\bqa && \mathcal{F} = - \frac{1}{\beta} \int_{-\infty}^{\infty} d \mu_{i}^{(0)} d t_{i}^{(0)} P^{(0)}[\mu_{i}^{(0)},t_{i}^{(0)}] \ln \int D c_{i \sigma} D \mu_{i}^{(0)} D t_{i}^{(0)} D \mu_{i}^{(1)} D t_{i}^{(1)} \nn && \delta(\mu_{i}^{(0)} - \mu + v_{i}) \delta(t_{i}^{(0)} - t_{i}) \delta\Big(\mu_{i}^{(1)} - \mu_{i}^{(0)} + \frac{2 t_{i}^{(0) 2}}{\mu_{i}^{(0)}}\Big) \delta\Big(t_{i}^{(1)} + \frac{t_{i}^{(0) 2}}{\mu_{i}^{(0)}}\Big) \nn && \exp\Big[ - \int_{0}^{\beta} d \tau \sum_{i = 1}^{N} \Big\{ c_{i\sigma}^{\dagger} (\partial_{\tau} - \mu_{i}^{(1)}) c_{i\sigma} - t_{i}^{(1)} (c_{i\sigma}^{\dagger} c_{i+1\sigma} + c_{i+1\sigma}^{\dagger} c_{i \sigma}) \Big\} + \frac{N_{s}}{2} \sum_{i = 1}^{N} \ln \Big( 1 + e^{ \beta \mu_{i}^{(0)} } \Big) \Big] , \label{1Loop_RG_FE_unrenor_distri} \eqa
where $\mu_{i}^{(1)} = \mu_{i}^{(0)} - \frac{2 t_{i}^{(0) 2}}{\mu_{i}^{(0)}}$ and $t_{i}^{(1)} = - \frac{t_{i}^{(0) 2}}{\mu_{i}^{(0)}}$ are renormalized chemical potential and hopping integral, respectively, introduced into two $\delta-$functions. This is the standard Kadanoff block-spin transformation. Here, we did not rewrite the distribution function of bare variables as that of renormalized ones yet.

%
%
Since the quenched-averaged free energy has to be also invariant under the renormalization group transformation, we obtain
\bqa && \int_{-\infty}^{\infty} d \mu_{i}^{(0)} d t_{i}^{(0)} P^{(0)}[\mu_{i}^{(0)},t_{i}^{(0)}] = \int_{-\infty}^{\infty} d \mu_{i}^{(1)} d t_{i}^{(1)} P^{(1)}[\mu_{i}^{(1)},t_{i}^{(1)}] . \eqa
It is straightforward to find the relation between $P^{(0)}[\mu_{i}^{(0)},t_{i}^{(0)}]$ and $P^{(1)}[\mu_{i}^{(1)},t_{i}^{(1)}]$, given by
\bqa && P^{(0)}[\mu_{i}^{(0)},t_{i}^{(0)}] = \Big( \frac{\partial \mu_{i}^{(1)}}{\partial \mu_{i}^{(0)}} \frac{\partial t_{i}^{(1)}}{\partial t_{i}^{(0)}} - \frac{\partial \mu_{i}^{(1)}}{\partial t_{i}^{(0)}} \frac{\partial t_{i}^{(1)}}{\partial \mu_{i}^{(0)}} \Big) P^{(1)}[\mu_{i}^{(1)},t_{i}^{(1)}] . \label{1Loop_RG_Distri} \eqa
Here, $\Big( \frac{\partial \mu_{i}^{(1)}}{\partial \mu_{i}^{(0)}} \frac{\partial t_{i}^{(1)}}{\partial t_{i}^{(0)}} - \frac{\partial \mu_{i}^{(1)}}{\partial t_{i}^{(0)}} \frac{\partial t_{i}^{(1)}}{\partial \mu_{i}^{(0)}} \Big)$ is the Jacobian factor to compensate the change of the `volume' integral.

Introducing Eq. (\ref{1Loop_RG_Distri}) into Eq. (\ref{1Loop_RG_FE_unrenor_distri}), we obtain an effective free energy with a quenched average as follows
\bqa && \mathcal{F} = - \frac{1}{\beta} \int_{-\infty}^{\infty} d \mu_{i}^{(1)} d t_{i}^{(1)} P^{(1)}[\mu_{i}^{(1)},t_{i}^{(1)}] \ln \int D c_{i \sigma} D \mu_{i}^{(0)} D t_{i}^{(0)} D \mu_{i}^{(1)} D t_{i}^{(1)} \nn && \delta(\mu_{i}^{(0)} - \mu + v_{i}) \delta(t_{i}^{(0)} - t_{i}) \delta\Big(\mu_{i}^{(1)} - \mu_{i}^{(0)} + \frac{2 t_{i}^{(0) 2}}{\mu_{i}^{(0)}}\Big) \delta\Big(t_{i}^{(1)} + \frac{t_{i}^{(0) 2}}{\mu_{i}^{(0)}}\Big) \nn && \exp\Big[ - \int_{0}^{\beta} d \tau \sum_{i = 1}^{N} \Big\{ c_{i\sigma}^{\dagger} (\partial_{\tau} - \mu_{i}^{(1)}) c_{i\sigma} - t_{i}^{(1)} (c_{i\sigma}^{\dagger} c_{i+1\sigma} + c_{i+1\sigma}^{\dagger} c_{i \sigma}) \Big\} + \frac{N_{s}}{2} \sum_{i = 1}^{N} \ln \Big( 1 + e^{ \beta \mu_{i}^{(0)} } \Big) \Big] . \label{1Loop_RG_FE} \eqa
This completes the first iteration step of the Kadanoff block-spin transformation.

Before going to further iterations of renormalization group transformations, we point out two types of approximations in the Kadanoff block-spin transformation. First of all, nonlocal terms are neglected in the resulting effective Lagrangian. The Wilsonian renormalization group transformation generates nonlocal terms inevitably. Within the present regularization scheme, nonlocal hopping terms along the time direction are given by the even-site electron propagator. Resorting to the gradient expansion for the time derivative in the even-site electron propagator, we keep the lowest order of $(\partial_{\tau} / \mu_{i})^{n}$ with $n = 0$. The other approximation is to assume local homogeneity in the hopping integral. In other words, we also neglect differences between neighbor hopping integrals in the sense of the gradient expansion for the spatial derivative.

\subsection{Recursive renormalization group transformations}

Now, we repeat the Kadanoff block-spin transformation in a recursive way. It is straightforward to generalize Eq. (\ref{1Loop_RG_FE}) as follows
\bqa && \mathcal{F} = - \frac{1}{\beta} \int_{-\infty}^{\infty} d \mu_{i}^{(f)} d t_{i}^{(f)} P^{(f)}[\mu_{i}^{(f)},t_{i}^{(f)}] \ln \int D c_{i \sigma} D \mu_{i}^{(0)} D t_{i}^{(0)} \delta(\mu_{i}^{(0)} - \mu + v_{i}) \delta(t_{i}^{(0)} - t_{i}) \nn && \int \Pi_{k = 1}^{f} D \mu_{i}^{(k)} D t_{i}^{(k)} \delta\Big(\mu_{i}^{(k)} - \mu_{i}^{(k-1)} + \frac{2 t_{i}^{(k-1) 2}}{\mu_{i}^{(k-1)}}\Big) \delta\Big(t_{i}^{(k)} + \frac{t_{i}^{(k-1) 2}}{\mu_{i}^{(k-1)}}\Big) \nn && \exp\Big[ - \int_{0}^{\beta} d \tau \sum_{i = 1}^{N} \Big\{ c_{i\sigma}^{\dagger} (\partial_{\tau} - \mu_{i}^{(f)}) c_{i\sigma} - t_{i}^{(f)} (c_{i\sigma}^{\dagger} c_{i+1\sigma} + c_{i+1\sigma}^{\dagger} c_{i \sigma}) \Big\} + \frac{N_{s}}{2} \sum_{k = 1}^{f} \sum_{i = 1}^{N} \ln \Big( 1 + e^{ \beta \mu_{i}^{(k-1)} } \Big) \Big] . \eqa
Two recursive equations of $\mu_{i}^{(k)} = \mu_{i}^{(k-1)} - \frac{2 t_{i}^{(k-1) 2}}{\mu_{i}^{(k-1)}}$ and $t_{i}^{(k)} = - \frac{t_{i}^{(k-1) 2}}{\mu_{i}^{(k-1)}}$ describe how $\mu_{i}^{(k-1)}$ and $t_{i}^{(k-1)}$ evolve to $\mu_{i}^{(k)}$ and $t_{i}^{(k)}$ through the renormalization group transformation, given by two $\delta-$functions. Fully renormalized chemical potential $\mu_{i}^{(f)}$ and hopping integral $t_{i}^{(f)}$ appear in the IR effective Lagrangian. The renormalized free energy functional
\bqa && F[\mu_{i}^{(f)},t_{i}^{(f)}] = - \frac{1}{\beta} \ln \int D c_{i \sigma} D \mu_{i}^{(0)} D t_{i}^{(0)} \delta(\mu_{i}^{(0)} - \mu + v_{i}) \delta(t_{i}^{(0)} - t_{i}) \nn && \int \Pi_{k = 1}^{f} D \mu_{i}^{(k)} D t_{i}^{(k)} \delta\Big(\mu_{i}^{(k)} - \mu_{i}^{(k-1)} + \frac{2 t_{i}^{(k-1) 2}}{\mu_{i}^{(k-1)}}\Big) \delta\Big(t_{i}^{(k)} + \frac{t_{i}^{(k-1) 2}}{\mu_{i}^{(k-1)}}\Big) \nn && \exp\Big[ - \int_{0}^{\beta} d \tau \sum_{i = 1}^{N} \Big\{ c_{i\sigma}^{\dagger} (\partial_{\tau} - \mu_{i}^{(f)}) c_{i\sigma} - t_{i}^{(f)} (c_{i\sigma}^{\dagger} c_{i+1\sigma} + c_{i+1\sigma}^{\dagger} c_{i \sigma}) \Big\} + \frac{N_{s}}{2} \sum_{k = 1}^{f} \sum_{i = 1}^{N} \ln \Big( 1 + e^{ \beta \mu_{i}^{(k-1)} } \Big) \Big] \nonumber \eqa
in a given disorder realization $(\mu_{i}^{(f)},t_{i}^{(f)})$ is quenched-averaged by the renormalized distribution function $P^{(f)}[\mu_{i}^{(f)},t_{i}^{(f)}]$ as
\bqa && \mathcal{F} = \int_{-\infty}^{\infty} d \mu_{i}^{(f)} d t_{i}^{(f)} P^{(f)}[\mu_{i}^{(f)},t_{i}^{(f)}] F[\mu_{i}^{(f)},t_{i}^{(f)}] . \nonumber \eqa
It is also straightforward to generalize Eq. (\ref{1Loop_RG_Distri}) to
\bqa && P^{(k-1)}[\mu_{i}^{(k-1)},t_{i}^{(k-1)}] = \Big( \frac{\partial \mu_{i}^{(k)}}{\partial \mu_{i}^{(k-1)}} \frac{\partial t_{i}^{(k)}}{\partial t_{i}^{(k-1)}} - \frac{\partial \mu_{i}^{(k)}}{\partial t_{i}^{(k-1)}} \frac{\partial t_{i}^{(k)}}{\partial \mu_{i}^{(k-1)}} \Big) P^{(k)}[\mu_{i}^{(k)},t_{i}^{(k)}] . \eqa

Another interesting idea is to replace the discrete index $(k)$ of the superscript with a continuum coordinate $z$ in the following way
\bqa && \mathcal{F} = - \frac{1}{\beta} \int_{-\infty}^{\infty} d \mu_{i}(z_{f}) d t_{i}(z_{f}) P[\mu_{i}(z_{f}),t_{i}(z_{f});z_{f}] \nn && \ln \int D c_{i \sigma} D \mu_{i}(z) D t_{i}(z) \delta(\mu_{i}(0) - \mu + v_{i}) \delta(t_{i}(0) - t_{i}) \delta\Big(\partial_{z} \mu_{i}(z) + \frac{2 t_{i}^{2}(z)}{\mu_{i}(z)}\Big) \delta\Big(\partial_{z} t_{i}(z) + t_{i}(z) + \frac{t_{i}^{2}(z)}{\mu_{i}(z)}\Big) \nn && \exp\Big[ - \int_{0}^{\beta} d \tau \sum_{i = 1}^{N} \Big\{ c_{i\sigma}^{\dagger} [\partial_{\tau} - \mu_{i}(z_{f})] c_{i\sigma} - t_{i}(z_{f}) (c_{i\sigma}^{\dagger} c_{i+1\sigma} + c_{i+1\sigma}^{\dagger} c_{i \sigma}) \Big\} + \frac{N_{s}}{2} \int_{0}^{z_{f}} d z \sum_{i = 1}^{N} \ln \Big( 1 + e^{\beta \mu_{i}(z)} \Big) \Big] . \eqa
Here, we consider $\mu_{i}^{(k)} - \mu_{i}^{(k-1)} = \partial_{z} \mu_{i}(z)$ and $z_{f} = f d z$, where $d z$ is an energy scale in each step of the renormalization group transformation. In this continuum expression, we do not introduce ghost fields explicitly into the effective action just for simplicity.
%
%

The power of this continuum-coordinate representation can be seen in the equation for the distribution function to satisfy, given by
\bqa && P[\mu_{i}(z),t_{i}(z);z] = \Big( \frac{\partial [\mu_{i}(z) + d z \partial_{z} \mu_{i}(z)]}{\partial \mu_{i}(z)} \frac{\partial [t_{i}(z) + d z \partial_{z} t_{i}(z)]}{\partial t_{i}(z)} \nn && - \frac{\partial [\mu_{i}(z) + d z \partial_{z} \mu_{i}(z)]}{\partial t_{i}(z)} \frac{\partial [t_{i}(z) + d z \partial_{z} t_{i}(z)]}{\partial \mu_{i}(z)} \Big) P[\mu_{i}(z) + d z \partial_{z} \mu_{i}(z),t_{i}(z) + d z \partial_{z} t_{i}(z);z + d z] . \eqa
As discussed in the previous section, the above relation is translated into the following differential equation
\bqa && \Big( \frac{\partial }{\partial z} + [\partial_{z} \mu_{i}(z)] \frac{\partial}{\partial \mu_{i}(z)} + [\partial_{z} t_{i}(z)] \frac{\partial}{\partial t_{i}(z)} \Big) \ln P[\mu_{i}(z),t_{i}(z);z] = - \frac{\partial [\partial_{z} \mu_{i}(z)]}{\partial \mu_{i}(z)} - \frac{\partial [\partial_{z} t_{i}(z)]}{\partial t_{i}(z)} , \eqa
which may be called the Callan-Symanzik equation for the distribution function. This completes our derivation for the present theoretical framework.

\subsection{Renormalization group flow of the distribution function}

The remaining task is to solve the renormalization group equations for all the coupling functions in a given disorder configuration at UV and find the distribution function for quenched averaging of the free energy. Introducing
\bqa && \mathcal{T}_{i}(z) = \frac{t_{i}(z)}{\mu_{i}(z)} \eqa
into the renormalization group equation for the hopping integral, we obtain
\bqa && \beta_{\mathcal{T}}[\mathcal{T}_{i}(z)] \equiv \partial_{z} \mathcal{T}_{i}(z) = \mathcal{T}_{i}(z) [2 \mathcal{T}_{i}(z) + 1] [\mathcal{T}_{i}(z) - 1] . \eqa
This renormalization group $\beta-$function shows three stable fixed points given by $\mathcal{T}_{i}^{*} = 0$ and $\mathcal{T}_{i}^{*} \rightarrow \pm \infty$, and two unstable ones given by $\mathcal{T}_{i}^{*} = 1$ and $\mathcal{T}_{i}^{*} = - \frac{1}{2}$. Below, we will discuss how these three stable fixed points in the clean limit evolve into novel ones in the disordered case. It is not difficult to solve this differential equation in a general case and find
\bqa && \Bigg( \frac{\mathcal{T}_{i}(z) - 1}{\mathcal{T}_{i}(0) - 1} \Bigg)^{2/3} \Bigg(\frac{\mathcal{T}_{i}(z) + 1/2}{\mathcal{T}_{i}(0) + 1/2}\Bigg)^{4/3} \Bigg( \frac{\mathcal{T}_{i}(0)}{\mathcal{T}_{i}(z)}\Bigg)^{2} = e^{2 z} . \eqa

The renormalization group flow for the chemical potential is given by
\bqa && \partial_{z} \ln \mu_{i}(z) = - 2 \mathcal{T}_{i}^{2}(z) , \eqa
resulting in
\bqa && \mu_{i}(z) = \mu_{i}(0) \exp\Big(- 2 \int_{0}^{z} d y \mathcal{T}_{i}^{2}(y) \Big) . \eqa
Near the insulating fixed point $\mathcal{T}_{i}^{*} = 0$, we have a fixed-point value of the chemical potential $\mu_{i}^{*}$, determined by its initial UV value.

Introducing the renormalization-group $\beta-$functions into the renormalization group flow of the distribution function, we obtain
\bqa && \Big\{ \frac{\partial }{\partial z} - \frac{2 t_{i}^{2}(z)}{\mu_{i}(z)} \frac{\partial}{\partial \mu_{i}(z)} - \Big(t_{i}(z) + \frac{t_{i}^{2}(z)}{\mu_{i}(z)}\Big) \frac{\partial}{\partial t_{i}(z)} \Big\} \ln P[\mu_{i}(z),t_{i}(z);z] = 1 + 2 \frac{t_{i}(z)}{\mu_{i}(z)} + 2 \frac{t_{i}^{2}(z)}{\mu_{i}^{2}(z)} . \eqa
To replace $t_{i}(z)$ with $\mathcal{T}_{i}(z)$, we have to consider
\bqa && \int_{-\infty}^{\infty} d \mu_{i}(z) d t_{i}(z) P[\mu_{i}(z),t_{i}(z);z] = \int_{-\infty}^{\infty} d \mu_{i}(z) d \mathcal{T}_{i}(z) \Big(\frac{\partial \mathcal{T}_{i}(z)}{\partial t_{i}(z)}\Big)^{-1} P[\mu_{i}(z),\mathcal{T}_{i}(z);z] . \eqa
As a result, the distribution function is changed from $P[\mu_{i}(z),t_{i}(z);z]$ to $\mu_{i}(z) P[\mu_{i}(z),\mathcal{T}_{i}(z);z]$. Based on this modification, we obtain the following renormalization group flow of the distribution function
\bqa && \Big\{ \frac{\partial }{\partial z} + \Big( 1 + \frac{1}{\mu_{i}(z)} \Big) [\partial_{z} \mu_{i}(z)] \frac{\partial}{\partial \mu_{i}(z)} + [\partial_{z} \mathcal{T}_{i}(z)] \frac{\partial}{\partial \mathcal{T}_{i}(z)} \Big\} \ln P[\mu_{i}(z),\mathcal{T}_{i}(z);z] = 1 + 2 \mathcal{T}_{i}(z) + 2 \mathcal{T}_{i}^{2}(z) . \eqa
Compared to the renormalization group flow in terms of original coupling functions, there appears a correction term given by $1 + \frac{1}{\mu_{i}(z)}$ in these newly introduced variables. Finally, we have
\bqa && \Big\{ \frac{\partial }{\partial z} - 2 \mathcal{T}_{i}^{2}(z) \Big( 1 + \mu_{i}(z) \Big) \frac{\partial}{\partial \mu_{i}(z)} + \mathcal{T}_{i}(z) [2 \mathcal{T}_{i}(z) + 1] [\mathcal{T}_{i}(z) - 1] \frac{\partial}{\partial \mathcal{T}_{i}(z)} \Big\} \ln P[\mu_{i}(z),\mathcal{T}_{i}(z);z] = 1 + 2 \mathcal{T}_{i}(z) + 2 \mathcal{T}_{i}^{2}(z) . \label{RG_Flow_Distribution_NIF} \nn \eqa

In this paper, we do not investigate the solution of this Callan-Symanzik equation for the distribution function in details. Instead, we focus on a near fixed-point solution. Here, we consider an insulating fixed point given by $\mathcal{T}_{i}^{*} = 0$ and $\mu_{i}^{*} = \mu_{i}(0)$, where the linearized renormalization group $\beta-$functions are
\bqa && \partial_{z} \delta \mathcal{T}_{i}(z) = - \delta \mathcal{T}_{i}(z) , ~~~~~ \partial_{z} \delta \mu_{i}(z) = 0 . \eqa
Introducing $\mathcal{T}_{i}(z) = \delta \mathcal{T}_{i}(z)$ and $\mu_{i}(z) = \mu_{i}(0) + \delta \mu_{i}(z)$ into Eq. (\ref{RG_Flow_Distribution_NIF}) and keeping all the terms up to the linear order for these variations, we obtain
\bqa && \Big\{ \frac{\partial }{\partial z} - \delta \mathcal{T}_{i}(z) \frac{\partial}{\partial \mathcal{T}_{i}(z)} \Big\} \ln P_{*}[\delta \mathcal{T}_{i}(z);z] = 1 + 2 \delta \mathcal{T}_{i}(z) . \eqa
This gives rise to
\bqa && P_{*}[\delta \mathcal{T}_{i}(z);z] = P_{*}[\mu_{i}(0),0;z_{0}] e^{z - z_{0}} \exp\Big\{- 2 [\delta \mathcal{T}_{i}(z) - \delta \mathcal{T}_{i}(z_{0})]\Big\} . \eqa

Considering the fixed point $\mathcal{T}_{i}^{*} = 0$, it is natural to interpret that this distribution function describes the localization physics of an Anderson insulating phase in one spatial dimension. This does not correspond to an infinite randomness fixed point. The renormalization group flow of the distribution function also allows an exponential tail in the Anderson localized phase. Recall that there are other fixed points $\mathcal{T}_{i}^{*} \rightarrow \pm \infty$. Here, we do not investigate the behavior of the distribution function near this fixed point, but we suspect that this clean fixed point may evolve into a random singlet state (spin language) or a critical metallic phase \cite{Strong_Disorder_RG}, identified with an infinite randomness fixed point genuinely. More careful studies have to be performed and compared with the strong-disorder renormalization group approach.

%
%
%
%
%
%

\section{A functional renormalization group transformation method for one-dimensional disordered interacting fermions}

\subsection{Recursive Kadanoff block-spin transformations}

It is straightforward to generalize the previous framework into the case of one-dimensional disordered interacting fermions. The quenched-averaged free energy is given by
\bqa && \mathcal{F} = - \frac{1}{\beta} \int_{-\infty}^{\infty} d \mu_{i} P(\mu_{i}) \ln \int D c_{i \sigma} \exp\Big[ - \int_{0}^{\beta} d \tau \sum_{i = 1}^{L} \Big\{ c_{i\sigma}^{\dagger} (\partial_{\tau} - \mu_{i}) c_{i\sigma} - t (c_{i\sigma}^{\dagger} c_{i+1\sigma} + c_{i+1\sigma}^{\dagger} c_{i\sigma}) + \frac{u}{2N} c_{i\sigma}^{\dagger} c_{i\sigma} c_{i\sigma'}^{\dagger} c_{i\sigma'} \Big\} \Big] , \nn \eqa
where density-density interactions are introduced. Here, we consider the chemical potential as a random variable, given by $\mu_{i} = \mu - v_{i}$.

In the presence of interactions, we perform the Hubbard-Stratonovich transformation to introduce a scalar potential dual to the density field, and repeat essentially the same renormalization group transformation as that of the previous section. As a result, we obtain the following partition function
\bqa && \mathcal{Z}[\mu_{i}] = \int D c_{i\sigma} D \varphi_{i}(z) D \mu_{i}(z) D t_{i}(z) D u_{i}(z) \delta\Big(\mu_{i}(0) - \mu + v_{i}\Big) \delta\Big(t_{i}(0) - t\Big) \delta\Big(u_{i}(0) - u\Big) \nn && \delta\Big\{ \partial_{z} \mu_{i}(z) + \frac{2 t_{i}^{2}(z) \mu_{i}(z)}{\mu_{i}^{2}(z) - u_{i}(z)} \Big\} \delta\Big\{ \partial_{z} t_{i}(z) + t_{i}(z) + \frac{t_{i}^{2}(z) \mu_{i}(z)}{\mu_{i}^{2}(z) - u_{i}(z)} - \frac{i}{2} \partial_{z} \varphi_{i}(z) \Big\} \nn && \delta\Big\{ \partial_{z} u_{i}(z) + u_{i}(z) - \frac{4 u_{i}(z) t_{i}^{4}(z)}{\mu_{i}^{2}(z)\Big(\mu_{i}^{2}(z) - u_{i}(z)\Big)} \Big\} \nn && \exp\Bigg[ - \int_{0}^{\beta} d \tau \sum_{i = 1}^{L} \Bigg\{ c_{i\sigma}^{\dagger} \Big( \partial_{\tau} - \mu_{i}(z_{f}) - i \varphi_{i}(z_{f}) \Big) c_{i\sigma} - t_{i}(z_{f}) (c_{i\sigma}^{\dagger} c_{i+1\sigma} + c_{i+1\sigma}^{\dagger} c_{i\sigma}) \Bigg\} \nn && - \int_{0}^{\beta} d \tau \sum_{i = 1}^{L} \frac{N}{4 u_{i}(0)} \varphi_{i}^{2}(0) - \int_{0}^{z_{f}} d z \int_{0}^{\beta} d \tau \sum_{i = 1}^{L} \Bigg\{ \frac{N}{4 u_{i}(z)} \Big( \partial_{z} \varphi_{i}(z) \Big)^{2} + \frac{1}{4} \ln \frac{N}{2} \Big(\frac{1}{u_{i}(z)} - \frac{1}{\mu_{i}^{2}(z)}\Big) \nn && + \frac{N}{4} \frac{u_{i}(z)}{\mu_{i}^{2}(z) - u_{i}(z)} \Bigg\} + \frac{N}{2} \int_{0}^{z_{f}} d z \sum_{i = 1}^{L} \mbox{tr}_{\tau\tau'} \ln \Big(\partial_{\tau} - \mu_{i}(z)\Big)_{\tau\tau'} \Bigg] \eqa
in a given disorder configuration $\mu_{i}(0)$. We point out that essentially the same renormalization group transformation has been performed in a one-dimensional $\phi^{4}-$type Landau-Ginzburg theory as an effective field theory of a transverse-field Ising model \cite{Einstein_Klein_Gordon_RG_Kim,RG_GR_Geometry_I_Kim,RG_GR_Geometry_II_Kim,Kitaev_Entanglement_Entropy_Kim,RG_Holography_First_Kim}. More detailed discussions in this partition function will be given below.

\subsection{Renormalization group flow of the distribution function}

The next procedure is to quenched-average the free energy $F[\mu_{i}] = - \frac{1}{\beta} \ln \mathcal{Z}[\mu_{i}]$ as follows
\bqa && \mathcal{F} = - \frac{1}{\beta} \int_{-\infty}^{\infty} d \mu_{i}(z_{f}) d t_{i}(z_{f}) d u_{i}(z_{f}) d \varphi_{i}(z_{f}) P[\mu_{i}(z_{f}),t_{i}(z_{f}),u_{i}(z_{f}),\varphi_{i}(z_{f});z_{f}] \ln \int D c_{i \sigma} D \varphi_{i}(z) D \mu_{i}(z) D t_{i}(z) D u_{i}(z) \nn && \delta\Big(\mu_{i}(0) - \mu + v_{i}\Big) \delta\Big(t_{i}(0) - t\Big) \delta\Big(u_{i}(0) - u\Big) \delta\Big\{ \partial_{z} \mu_{i}(z) + \frac{2 t_{i}^{2}(z) \mu_{i}(z)}{\mu_{i}^{2}(z) - u_{i}(z)} \Big\} \delta\Big\{ \partial_{z} t_{i}(z) + t_{i}(z) + \frac{t_{i}^{2}(z) \mu_{i}(z)}{\mu_{i}^{2}(z) - u_{i}(z)} - \frac{i}{2} \partial_{z} \varphi_{i}(z) \Big\} \nn && \delta\Big\{ \partial_{z} u_{i}(z) + u_{i}(z) - \frac{4 u_{i}(z) t_{i}^{4}(z)}{\mu_{i}^{2}(z) [\mu_{i}^{2}(z) - u_{i}(z)]} \Big\} \exp\Big[ - \int_{0}^{\beta} d \tau \sum_{i = 1}^{N} \Big\{ c_{i\sigma}^{\dagger} [\partial_{\tau} - \mu_{i}(z_{f}) - i \varphi_{i}(z_{f})] c_{i\sigma} \nn && - t_{i}(z_{f}) (c_{i\sigma}^{\dagger} c_{i+1\sigma} + c_{i+1\sigma}^{\dagger} c_{i \sigma}) \Big\} - N_{s} \int_{0}^{\beta} d \tau \sum_{i = 1}^{N} \frac{1}{4 u} \varphi_{i}^{2}(0) - N_{s} \int_{0}^{z_{f}} d z \int_{0}^{\beta} d \tau \sum_{i = 1}^{N} \Big\{ \frac{1}{4 u_{i}(z)} [\partial_{z} \varphi_{i}(z)]^{2} \nn && + \frac{1}{4} \frac{u_{i}(z)}{\mu_{i}^{2}(z) - u_{i}(z)} - \frac{1}{2 \beta} \ln \Big( 1 + e^{\beta \mu_{i}(z)} \Big) \Big\} \Big] . \label{Quenched_Average_FE_Interaction} \eqa
As discussed before, the essential point is to rewrite the distribution function of unrenormalized UV variables as that of IR renormalized ones, based on the following equation
\bqa && P[\mu_{i}(z),t_{i}(z),u_{i}(z),\varphi_{i}(z);z] = \mbox{Det} \begin{pmatrix} 1 + d z \frac{\partial [\partial_{z} \mu_{i}(z)]}{\partial \mu_{i}(z)} & d z \frac{\partial [\partial_{z} \mu_{i}(z)]}{\partial t_{i}(z)} & d z \frac{\partial [\partial_{z} \mu_{i}(z)]}{\partial u_{i}(z)} & d z \frac{\partial [\partial_{z} \mu_{i}(z)]}{\partial \varphi_{i}(z)} \\ d z \frac{\partial [\partial_{z} t_{i}(z)]}{\partial \mu_{i}(z)} & 1 + d z \frac{\partial [\partial_{z} t_{i}(z)]}{\partial t_{i}(z)} & d z \frac{\partial [\partial_{z} t_{i}(z)]}{\partial u_{i}(z)} & d z \frac{\partial [\partial_{z} t_{i}(z)]}{\partial \varphi_{i}(z)} \\ d z \frac{\partial [\partial_{z} u_{i}(z)]}{\partial \mu_{i}(z)} & d z \frac{\partial [\partial_{z} u_{i}(z)]}{\partial t_{i}(z)} & 1 + d z \frac{\partial [\partial_{z} u_{i}(z)]}{\partial u_{i}(z)} & d z \frac{\partial [\partial_{z} u_{i}(z)]}{\partial \varphi_{i}(z)} \\ d z \frac{\partial [\partial_{z} \varphi_{i}(z)]}{\partial \mu_{i}(z)} & d z \frac{\partial [\partial_{z} \varphi_{i}(z)]}{\partial t_{i}(z)} & d z \frac{\partial [\partial_{z} \varphi_{i}(z)]}{\partial u_{i}(z)} & 1 + d z \frac{\partial [\partial_{z} \varphi_{i}(z)]}{\partial \varphi_{i}(z)} \end{pmatrix} \nn && P[\mu_{i}(z) + d z \partial_{z} \mu_{i}(z),t_{i}(z) + d z \partial_{z} t_{i}(z),u_{i}(z) + d z \partial_{z} u_{i}(z),\varphi_{i}(z) + d z \partial_{z} \varphi_{i}(z);z + d z] . \eqa
It is straightforward to rewrite this expression as the below Callan-Symanzik equation for the distribution function
\bqa && \Big(\frac{\partial}{\partial z} + \partial_{z} \mu_{i}(z) \frac{\partial}{\partial \mu_{i}(z)} + \partial_{z} t_{i}(z) \frac{\partial}{\partial t_{i}(z)} + \partial_{z} u_{i}(z) \frac{\partial}{\partial u_{i}(z)} + \partial_{z} \varphi_{i}(z) \frac{\partial}{\partial \varphi_{i}(z)} \Big) \ln P[\mu_{i}(z),t_{i}(z),u_{i}(z),\varphi_{i}(z);z] \nn && = - \frac{\partial [\partial_{z} \mu_{i}(z)]}{\partial \mu_{i}(z)} - \frac{\partial [\partial_{z} t_{i}(z)]}{\partial t_{i}(z)} - \frac{\partial [\partial_{z} u_{i}(z)]}{\partial u_{i}(z)} - \frac{\partial [\partial_{z} \varphi_{i}(z)]}{\partial \varphi_{i}(z)} . \eqa

Renormalization group $\beta-$functions for three coupling functions are encoded into three $\delta-$functions of the quenched-averaged free energy, given by
\bqa && \partial_{z} \mu_{i}(z) = - \frac{2 t_{i}^{2}(z) \mu_{i}(z)}{\mu_{i}^{2}(z) - u_{i}(z)} , \\ && \partial_{z} t_{i}(z) = - t_{i}(z) - \frac{t_{i}^{2}(z) \mu_{i}(z)}{\mu_{i}^{2}(z) - u_{i}(z)} + \frac{i}{2} \partial_{z} \varphi_{i}(z) , \\ && \partial_{z} u_{i}(z) = - u_{i}(z) + \frac{4 u_{i}(z) t_{i}^{4}(z)}{\mu_{i}^{2}(z)\Big(\mu_{i}^{2}(z) - u_{i}(z)\Big)} . \label{RG_Flows_Coupling_Functions_With_Interactions} \eqa
These first-order differential equations require three integral constants, given by their UV data, $\mu_{i}(0) = \mu - v_{i}$, $t_{i}(0) = t$, and $u_{i}(0) = u$.

The renormalization group flow for the scalar field $\varphi_{i}(z)$ dual to the density field at UV is given by minimization of the free energy with respect to $\varphi_{i}(z)$, resulting in
\bqa && - \partial_{z}^{2} \varphi_{i}(z) + [\partial_{z} \ln u_{i}(z)] [\partial_{z} \varphi_{i}(z)] = 0 . \eqa
This second-order differential equation requests two boundary conditions. The UV boundary condition can be determined by
\bqa && \partial_{z} \Big(\frac{\partial \mathcal{L}_{UV}}{\partial [\partial_{z} \varphi_{i}(z)]}\Big)_{z = 0} - \frac{\partial \mathcal{L}_{UV}}{\partial \varphi_{i}(z)} \Big|_{z = 0} = 0 , \eqa
where the UV boundary Lagrangian is
\bqa && \mathcal{L}_{UV} = \frac{N_{s}}{4 u_{i}(0)} \varphi_{i}^{2}(0) - \frac{N_{s}}{4 u_{i}(0)} \varphi_{i}(0) [\partial_{z} \varphi_{i}(z)]_{z = 0} , \eqa
derived from the free energy Eq. (\ref{Quenched_Average_FE_Interaction}). As a result, the UV boundary value of the scalar potential is $\varphi_{i}(0) = 0$. The IR boundary condition can be also derived from the free energy Eq. (\ref{Quenched_Average_FE_Interaction}). The IR boundary effective Lagrangian to include $\varphi_{i}(z_{f})$ is
\bqa && \mathcal{L}_{IR} = - i \varphi_{i}(z_{f}) \Big\langle c_{i\sigma}^{\dagger} c_{i\sigma} \Big\rangle + \frac{N_{s}}{4 u_{i}(z_{f})} \varphi_{i}(z_{f}) [\partial_{z} \varphi_{i}(z)]_{z = z_{f}} . \eqa
The Euler-Lagrange equation gives rise to the IR boundary condition for the IR scalar field $\varphi_{i}(z_{f})$,
\bqa && \partial_{z} \Big( \frac{1}{4 u_{i}(z)} \varphi_{i}(z) \Big)_{z = z_{f}} + i \Big\langle \frac{1}{N_{s}} \sum_{\sigma = 1}^{N_{s}} c_{i\sigma}^{\dagger} c_{i\sigma} \Big\rangle = 0 . \eqa
Here, $\Big\langle \frac{1}{N_{s}} \sum_{\sigma = 1}^{N_{s}} c_{i\sigma}^{\dagger} c_{i\sigma} \Big\rangle$ represents the ensemble average for the IR effective action
\bqa && S_{IR} = \int_{0}^{\beta} d \tau \sum_{i = 1}^{N} \Big\{ c_{i\sigma}^{\dagger} [\partial_{\tau} - \mu_{i}(z_{f}) - i \varphi_{i}(z_{f})] c_{i\sigma} - t_{i}(z_{f}) (c_{i\sigma}^{\dagger} c_{i+1\sigma} + c_{i+1\sigma}^{\dagger} c_{i \sigma}) \Big\} . \eqa
Now, all the ingredients of the renormalization group flow for the distribution function are completely determined.

\subsection{General structure near a fixed point}

Solving all these renormalization group equations for three coupling functions and one collective dual scalar field is quite involved to be beyond the scope of the present study. We suspect that our self-consistent non-perturbative renormalization-group improved mean-field theory would reproduce the Luttinger-liquid physics \cite{Luttinger_Liquid} in the clean limit. Following the discussion on the general structure of this field theory, it is natural to expect that the distribution function follow a power-law behavior near a disorder-modified conformally invariant fixed point, where the power-law critical exponent are given by those of the Luttinger-liquid-type fixed point in a given disorder configuration. Although we leave this investigation for a future study, we discuss a general structure near a fixed point.

It is straightforward to solve the equation of motion for the dual scalar field and to obtain
\bqa && \varphi_{i}(z) = \varphi_{i}(0) + \frac{[\partial_{z} \varphi_{i}(z)]_{z = 0}}{u_{i}(0)} \int_{0}^{z} d y u_{i}(y) . \eqa
Applying the UV boundary condition to this solution, we obtain $\varphi_{i}(0) = 0$ as mentioned before. One may rewrite the solution in the following way
\bqa && \varphi_{i}(z) = \frac{\varphi_{i}(z_{f})}{\int_{0}^{z_{f}} d y u_{i}(y)} \int_{0}^{z} d y u_{i}(y) , \eqa
where $\varphi_{i}(z_{f})$ is determined by the IR boundary condition.

Now, we consider the renormalization group flows for the coupling functions. Introducing
\bqa && \mathcal{T}_{i}(z) \equiv \frac{t_{i}(z)}{\mu_{i}(z)} , ~~~~~ \mathcal{U}_{i}(z) \equiv \frac{u_{i}(z)}{\mu_{i}^{2}(z)} \eqa
into Eq. (\ref{RG_Flows_Coupling_Functions_With_Interactions}), we obtain
\bqa && \partial_{z} \mathcal{T}_{i}(z) = - \mathcal{T}_{i}(z) - \frac{\mathcal{T}_{i}^{2}(z)}{1 - \mathcal{U}_{i}(z)} + \frac{2 \mathcal{T}_{i}^{3}(z) }{1 - \mathcal{U}_{i}(z)} + \frac{i \varphi_{i}(z_{f})}{2 \int_{0}^{z_{f}} d y u_{i}(y)} \mu_{i}^{2}(z) \mathcal{U}_{i}(z) , \nn && \partial_{z} \mathcal{U}_{i}(z) = - \mathcal{U}_{i}(z) + \frac{4 \mathcal{U}_{i}(z) \mathcal{T}_{i}^{2}(z)}{1 - \mathcal{U}_{i}(z)} + \frac{4 \mathcal{U}_{i}(z) \mathcal{T}_{i}^{4}(z)}{1 - \mathcal{U}_{i}(z)} , \nn && \partial_{z} \mu_{i}(z) = - \frac{2 \mathcal{T}_{i}^{2}(z)}{1 - \mathcal{U}_{i}(z)} \mu_{i}(z) . \eqa
Considering this change of coupling functions into the renormalization group flow of the distribution function, we find
\bqa && \Big\{\frac{\partial}{\partial z} + \Big(1 + \frac{3}{\mu_{i}(z)}\Big) \partial_{z} \mu_{i}(z) \frac{\partial}{\partial \mu_{i}(z)} + \partial_{z} \mathcal{T}_{i}(z) \frac{\partial}{\partial \mathcal{T}_{i}(z)} + \partial_{z} \mathcal{U}_{i}(z) \frac{\partial}{\partial \mathcal{U}_{i}(z)} + \partial_{z} \varphi_{i}(z) \frac{\partial}{\partial \varphi_{i}(z)} \Big\} \ln P[\mu_{i}(z),\mathcal{T}_{i}(z),\mathcal{U}_{i}(z),\varphi_{i}(z);z] \nn && = - \frac{\partial [\partial_{z} \mu_{i}(z)]}{\partial \mu_{i}(z)} - \frac{\partial [\partial_{z} \mathcal{T}_{i}(z)]}{\partial \mathcal{T}_{i}(z)} - \frac{\partial [\partial_{z} \mathcal{U}_{i}(z)]}{\partial \mathcal{U}_{i}(z)}- \frac{\partial [\partial_{z} \varphi_{i}(z)]}{\partial \varphi_{i}(z)} , \label{RG_Flow_Distribuion_Ft_With_Interaction} \eqa
where there appears a correction term given by $1 + \frac{3}{\mu_{i}(z)}$. Here, the number $3$ arises from $P[\mu_{i}(z),t_{i}(z),u_{i}(z),\varphi_{i}(z);z]$ $\longrightarrow$ $\Big(\frac{\partial \mathcal{T}_{i}(z)}{\partial t_{i}(z)}\Big)^{-1} \Big(\frac{\partial \mathcal{U}_{i}(z)}{\partial u_{i}(z)}\Big)^{-1} P[\mu_{i}(z),t_{i}(z),u_{i}(z),\varphi_{i}(z);z] = \mu_{i}^{3}(z) P[\mu_{i}(z),t_{i}(z),u_{i}(z),\varphi_{i}(z);z]$.

To determine a fixed point, we first consider $\partial_{z} \mathcal{U}_{i}(z) \Big|_{z \rightarrow \infty} = 0$. Then, we obtain either (i) $\mathcal{U}_{i}^{*} = 0$ or (ii) $\mathcal{U}_{i}^{*} = 1 - 4 \mathcal{T}_{i}^{* 2}(z) - 4 \mathcal{T}_{i}^{* 4}$. (i) is nothing but the non-interacting fixed point, where $\partial_{z} \mathcal{T}_{i}(z) \Big|_{z \rightarrow \infty} = 0$ with $\mathcal{U}_{i}^{*} = 0$ gives exactly the fixed-point structure discussed in the previous section. On the other hand, the fixed point (ii) leads to $4 \mathcal{T}_{i}^{* 3} + 2 \mathcal{T}_{i}^{* 2} + 1 = 0$, where $\mu_{i}^{*} = 0$ is taken into account from the renormalization group flow of the chemical potential. Interestingly, the non-interacting fixed point $\mathcal{T}_{i}^{*} = 0$ does not satisfy this equation. Instead, the fixed-point value is given by a complex number, rather unexpected.

Near a fixed point, the renormalization group flows of effective hopping and interaction parameters are given by
\bqa && \partial_{z} \delta \mathcal{T}_{i}(z) = \nu_{\mathcal{T}\mathcal{T}} \delta \mathcal{T}_{i}(z) + \nu_{\mathcal{T}\mathcal{U}} \delta \mathcal{U}_{i}(z) , \nn && \partial_{z} \delta \mathcal{U}_{i}(z) = \nu_{\mathcal{U}\mathcal{T}} \delta \mathcal{T}_{i}(z) + \nu_{\mathcal{U}\mathcal{U}} \delta \mathcal{U}_{i}(z) . \eqa
Here, critical exponents are as a function of the fixed-point values of $\mathcal{T}_{i}^{*}$ and $\mathcal{U}_{i}^{*}$, respectively. Introducing these linearized equations into Eq. (\ref{RG_Flow_Distribuion_Ft_With_Interaction}), we obtain
\bqa && \Big\{\frac{\partial}{\partial z} + \Big(\nu_{\mathcal{T}\mathcal{T}} \delta \mathcal{T}_{i}(z) + \nu_{\mathcal{T}\mathcal{U}} \delta \mathcal{U}_{i}(z)\Big) \frac{\partial}{\partial \delta \mathcal{T}_{i}(z)} + \Big(\nu_{\mathcal{U}\mathcal{T}} \delta \mathcal{T}_{i}(z) + \nu_{\mathcal{U}\mathcal{U}} \delta \mathcal{U}_{i}(z)\Big) \frac{\partial}{\partial \delta \mathcal{U}_{i}(z)} \Big\} \ln P_{*}[\delta \mathcal{T}_{i}(z),\delta \mathcal{U}_{i}(z);z] \nn && = - \nu_{\mu\mu} - \nu_{\mathcal{T}\mathcal{T}} - \nu_{\mathcal{U}\mathcal{U}} - \nu_{\varphi\varphi} . \eqa
One can verify that both contributions of $\Big(1 + \frac{3}{\mu_{i}(z)}\Big) \partial_{z} \mu_{i}(z) \frac{\partial}{\partial \mu_{i}(z)}$ and $\partial_{z} \varphi_{i}(z) \frac{\partial}{\partial \varphi_{i}(z)}$ do not appear near the fixed point of $\mu_{i}^{*} = 0$. This differential equation leads to a power-law distribution function, generally.

\section{Holographic dual effective field theory for disordered interacting electrons}

Finally, we generalize the previous theoretical framework to that in higher spatial dimensions than one. Resorting to the AdS$_{D+1}$/CFT$_{D}$ duality conjecture, we propose the following quenched-averaged free energy
\bqa && \mathcal{F} = \lim_{z_{f} \rightarrow \infty} \int D g_{\mu\nu}(x,z_{f}) P[g_{\mu\nu}(x,z_{f})] F[g_{\mu\nu}(x,z_{f})] , \eqa
where the effective free-energy functional in terms of the IR boundary metric tensor $g_{\mu\nu}(x,z_{f})$ is
\bqa && F[g_{\mu\nu}(x,z_{f})] = - \frac{1}{\beta} \ln \int D g_{\mu\nu}(x,z) D \pi^{\mu\nu}(x,z) \exp\Big[ - N_{c}^{2} \int_{0}^{z_{f}} d z \int d^{D} x \Big\{ \pi^{\mu\nu}(x,z) \partial_{z} g_{\mu\nu}(x,z) \nn && + \frac{\lambda}{2} \frac{1}{\sqrt{g(x,z)}} \pi^{\mu\nu}(x,z) \mathcal{G}_{\mu\nu\rho\gamma}(x,z) \pi^{\rho\gamma}(x,z) + \frac{1}{2 \kappa} \sqrt{g(x,z)} \Big( R(x,z) - 2 \Lambda \Big) \Big\} \Big] . \eqa
Here, $P[g_{\mu\nu}(x,z_{f})]$ is a renormalized distribution function, governed by
%
%
\bqa && \Big(\frac{\partial}{\partial z} + [\partial_{z} g_{\tau\tau}(x,z)] \frac{\partial}{\partial g_{\tau\tau}(x,z)} + [\partial_{z} g_{ij}(x,z)] \frac{\partial}{\partial g_{ij}(x,z)} \Big) \ln P[g_{\tau\tau}(x,z),g_{ij}(x,z);z] \nn && = - \frac{\partial [\partial_{z} g_{\tau\tau}(x,z)]}{\partial g_{\tau\tau}(x,z)} - \frac{\partial [\partial_{z} g_{ij}(x,z)]}{\partial g_{ij}(x,z)} . \eqa
The Jacobian factor is expressed symbolically, to be clarified below. In the dual gravity action, we consider the gaussian normal coordinate system in the Arnowitt-Deser-Misner (ADM) decomposition \cite{ADM_Hamiltonian_Formulation}, given by
\bqa && d s^{2} = \Big( \mathcal{N}^{2}(x,z) + \mathcal{N}_{\mu}(x,z) \mathcal{N}^{\mu}(x,z) \Big) d z^{2} + 2 \mathcal{N}_{\mu}(x,z) d x^{\mu} d z + g_{\mu\nu}(x,z) d x^{\mu} d x^{\nu} , \eqa
where the gauge fixing condition for the lapse function $\mathcal{N}(x,z) = 1$ and the shift vector $\mathcal{N}_{\mu}(x,z) = 0$ is taken into account \cite{GR_Textbook}. In addition, we consider the case of finite temperatures, meaning that the time circle $S^{1}$ is assumed with the periodicity $\beta$. $\mu, \nu$ cover from $\tau$ to $i, j = 1, ..., D-1$. In other words, $g_{\mu\nu}(x,z)$ is a $D-$dimensional metric tensor at a given slice $z$. $R(x,z)$ is the corresponding $D-$dimensional Ricci scalar. $\pi^{\mu\nu}(x,z)$ is the canonically conjugate field to the metric tensor $g_{\mu\nu}(x,z)$. $\mathcal{G}_{\mu\nu\rho\gamma}(x,z) \equiv g_{\mu\rho}(x,z) g_{\nu\gamma}(x,z) - \frac{1}{D-1} g_{\mu\nu}(x,z) g_{\rho\gamma}(x,z)$ is de Witt supermetric \cite{DeWitt_Metric}, taking into account transverseness. $N_{c}$ represents the color index. Taking the $z_{f} \rightarrow \infty$ limit, the dual gravity action reduces into the conventional holographic description as a gauge-fixed version.

The Euler-Lagrange equation of motion reads
\begin{equation}
  \partial_{z}^{2} g_{\mu\nu}(x,z) - g_{\mu\nu}(x,z) g^{\rho\sigma}(x,z) \partial_{z}^{2} g_{\rho\sigma}(x,z) = \frac{\lambda}{2 \kappa \sqrt{g(x,z)}} \Big(R_{\mu\nu}(x,z) - \frac{1}{2} R(x,z) g_{\mu\nu}(x,z) + \Lambda g_{\mu\nu}(x,z) \Big) .
\end{equation}
%
%
%
%
Both UV and IR boundary conditions are given according to the conventional dual holographic dictionary. Since this equation of motion is essentially identical to that of the dual gravity action (with introduction of both the lapse function and the shift vector), the background geometry is given by an AdS black hole.

Considering small fluctuations as
\bqa && g_{\mu\nu}(x,z) = g_{\mu\nu}^{BH}(x,z) + h_{\mu\nu}(x,z) \eqa
in the background black hole solution, one can find a linearized equation of motion for weak metric perturbations as follows
\begin{equation}
\begin{aligned}
   & \frac{g^{BH}_{\mu\alpha} g^{BH}_{\nu\beta}}{2\sqrt{g_{BH}}} \partial_{z} \bigg[\frac{\sqrt{g_{BH}}}{\lambda} \Big(g_{BH}^{\rho\sigma} h_{\rho\sigma} \mathcal{G}_{BH}^{\alpha\beta\gamma\delta} \big(\partial_{z} g^{BH}_{\gamma\delta}\big)
  +2\big(\delta_{h} \mathcal{G}_{BH}^{\alpha\beta\gamma\delta} \big) \big(\partial_{z} g^{BH}_{\gamma\delta}\big)
  +2 \mathcal{G}_{BH}^{\alpha\beta\gamma\delta} \big(\partial_{z} h_{\gamma\delta}\big)\Big) \bigg]
  \\
  &+ \frac{2h_{(\mu|\alpha|}g^{BH}_{\nu)\beta}}{\sqrt{g_{BH}}} \partial_{z}\bigg[\frac{\sqrt{g_{BH}}}{\lambda} \mathcal{G}_{BH}^{\alpha\beta\gamma\delta} \big(\partial_{z} g^{BH}_{\gamma\delta}\big)\bigg]
  - \frac{g^{BH}_{\mu\nu} \partial_{z} g^{BH}_{\alpha\beta}}{2\lambda} \big(\partial_{z} g^{BH}_{\alpha\beta}\big) \mathcal{G}_{BH}^{\alpha\beta\gamma\delta} \big(\partial_{z} h_{\gamma\delta}\big)
  \\
  &- \frac{g^{BH}_{\mu\nu}}{4\lambda} \big(\partial_{z} g^{BH}_{\alpha\beta}\big) \big(\delta_{h} \mathcal{G}_{BH}^{\alpha\beta\gamma\delta} \big) \big(\partial_{z} g^{BH}_{\gamma\delta}\big)
   -\frac{h_{\mu\nu}}{4\lambda} \big(\partial_{z} g^{BH}_{\alpha\beta}\big) \mathcal{G}_{BH}^{\alpha\beta\gamma\delta} \big(\partial_{z} g^{BH}_{\gamma\delta}\big)
   -\frac{h^{\rho\sigma}}{\lambda}\Big(\big(\partial_{z} g^{BH}_{\mu\rho}\big) \big(\partial_{z} g^{BH}_{\nu\sigma}\big) - \big(\partial_{z} g^{BH}_{\mu\nu}\big) \big(\partial_{z} g^{BH}_{\rho\sigma}\big)\Big)
   \\
   &+\frac{g_{BH}^{\rho\sigma}}{\lambda} \Big(
    \big(\partial_{z} g^{BH}_{\mu\rho}\big) \big(\partial_{z} h_{\nu\rho}\big)
   + \big(\partial_{z} h_{\mu\rho}\big) \big(\partial_{z} g^{BH}_{\nu\rho}\big)
   - \big(\partial_{z} g^{BH}_{\mu\nu}\big) \big(\partial_{z} h_{\rho\sigma}\big)
   -\big(\partial_{z} h_{\mu\nu}\big) \big(\partial_{z} g^{BH}_{\rho\sigma}\big) \Big)
   \\
   &+\frac{1}{4\kappa} \bigg[2h_{(\mu}{ }^{\alpha} R^{BH}_{\nu) \alpha} - 2h^{\alpha \beta} R^{BH}_{\mu \alpha \nu \beta} + \nabla_{BH}^{\rho} \Big(\nabla^{BH}_{\mu} h_{\rho\nu} {+} \nabla^{BH}_{\nu} h_{\rho\mu} {-} \nabla^{BH}_{\rho} h_{\mu\nu}\Big) - \nabla^{BH}_{\nu} \nabla^{BH}_{\mu} h^{\rho}{}_{\rho}
   \\
   &+ g^{BH}_{\mu\nu}\Big(h^{\rho\sigma} R^{BH}_{\rho\sigma} {-} \nabla^{BH}_{\rho} \nabla^{BH}_{\sigma} h^{\rho\sigma} {+} \nabla_{BH}^{\rho} \nabla^{BH}_{\rho} h^{\sigma}{}_{\sigma}\Big)- h_{\mu\nu} \big(R_{BH}-2\Lambda\big)\bigg]=0 .
\end{aligned}
\end{equation}
Here, we introduced the following notation for simplicity
\begin{equation}
  \delta_{h} \mathcal{G}_{BH}^{\alpha\beta\gamma\delta} = - \big(h^{\alpha(\gamma} g_{BH}^{\delta)\beta} + h^{\beta(\gamma} g_{BH}^{\delta)\alpha} - h^{\alpha\beta} g_{BH}^{\gamma\delta} - g_{BH}^{\alpha\beta} h^{\gamma\delta} \big) .
\end{equation}
Symmetrization for either super- or subscript symbols are easily understood and conventionally utilized. Accordingly, we can find the renormalization group flow for the distribution function near the conformally invariant fixed point as follows
\bqa && \Big(\frac{\partial}{\partial z} + [\partial_{z} h_{\tau\tau}(x,z)] \frac{\partial}{\partial h_{\tau\tau}(x,z)} + [\partial_{z} h_{x x}(x,z)] \frac{\partial}{\partial h_{x x}(x,z)} \Big) \ln P_{*}[h_{\tau\tau}(x,z), h_{x x}(x,z);z] \nn && = - \frac{\partial [\partial_{z} h_{\tau\tau}(x,z)]}{\partial h_{\tau\tau}(x,z)} - \frac{\partial [\partial_{z} h_{x x}(x,z)]}{\partial h_{x x}(x,z)} . \eqa
Here, the renormalization group flow for the weak perturbation of the metric tensor is taken into account, and only the case of one spatial dimension is considered for simplicity.

%
%

Based on the solution of the weak metric fluctuation near the AdS black hole background geometry, we obtain
\bqa && \partial_{z} \begin{pmatrix} h_{\tau\tau}(x,z) \\ h_{xx}(x,z) \end{pmatrix} = \begin{pmatrix} \nu_{\tau\tau}^{\tau\tau} & \nu_{\tau\tau}^{xx} \\ \nu_{xx}^{\tau\tau} & \nu_{xx}^{xx} \end{pmatrix} \begin{pmatrix} h_{\tau\tau}(x,z) \\ h_{xx}(x,z) \end{pmatrix} , \eqa
where the physical meaning of the critical-exponent matrix is clear. It is straightforward to diagonalize the critical-exponent matrix as
\bqa && \partial_{z} \begin{pmatrix} \mathcal{H}_{\tau\tau}(x,z) \\ \mathcal{H}_{xx}(x,z) \end{pmatrix} = \begin{pmatrix} \eta_{\tau\tau} & 0 \\ 0 & \eta_{xx} \end{pmatrix} \begin{pmatrix} \mathcal{H}_{\tau\tau}(x,z) \\ \mathcal{H}_{xx}(x,z) \end{pmatrix} . \eqa
As a result, we obtain
\bqa && \Big(\frac{\partial}{\partial z} + \eta_{\tau\tau} \frac{\partial}{\partial \ln \mathcal{H}_{\tau\tau}(x,z)} + \eta_{xx} \frac{\partial}{\partial \ln \mathcal{H}_{xx}(x,z)} \Big) \ln P_{*}[\mathcal{H}_{\tau\tau}(x,z), \mathcal{H}_{x x}(x,z);z] = - \nu_{\tau\tau}^{\tau\tau} - \nu_{xx}^{xx} . \eqa
This linearized Callan-Symanzik equation gives the following solution for the distribution function near the fixed point
%
%
\bqa && P_{*}[\mathcal{H}_{\tau\tau}(x,z), \mathcal{H}_{x x}(x,z);z] = \frac{\mathcal{C}^{*}}{[\mathcal{H}_{\tau\tau}(x,z)]^{\frac{\nu_{\tau\tau}^{\tau\tau} + \nu_{xx}^{xx}}{\eta_{\tau\tau}}} [\mathcal{H}_{xx}(x,z)]^{\frac{\nu_{\tau\tau}^{\tau\tau} + \nu_{xx}^{xx}}{\eta_{xx}}}} . \eqa
This power-law solution is determined by the critical-exponent matrix in the absence of randomness.

%
%

%
%

Resorting to this power-law distribution function of the metric tensor near the AdS$_{D+1}$ black hole geometry, we find an effective free energy as follows. First, we obtain an effective on-shell action from the dual holographic effective field theory in the large $N_{c}$ limit, given by
\bqa && F[g_{\mu\nu}(\tau,x,z_{f}), g_{\mu\nu}(\tau,x,\varepsilon)] = \frac{N_{c}^{2}}{\beta} \int_{0}^{\beta} d \tau \int d^{D-1} x \Big\{ \pi^{\mu\nu}(\tau,x,z_{f}) g_{\mu\nu}(\tau,x,z_{f}) - \pi^{\mu\nu}(\tau,x,\varepsilon) g_{\mu\nu}(\tau,x,\varepsilon) \Big\} . \eqa
Here, the Euclidean time circle was explicitly shown and $\varepsilon$ is the UV boundary. The IR boundary canonical momentum tensor corresponds to the IR boundary energy-momentum tensor \cite{Brown_York_Boundary_Term}, given by extrinsic curvatures \cite{GR_Textbook} at $z = z_{f}$ as follows
\bqa && \pi_{\mu\nu}(\tau,x,z_{f}) = \frac{1}{2 \kappa_{f}} \Big( R_{\mu\nu}(\tau,x,z_{f}) - \frac{1}{2} R(\tau,x,z_{f}) g_{\mu\nu}(\tau,x,z_{f}) \Big) . \eqa
As a result, we obtain the following expression for the on-shell effective action
\bqa && F[g_{\mu\nu}(\tau,x,z_{f}), g_{\mu\nu}(\tau,x,\varepsilon)] = \frac{(D-2) N_{c}^{2}}{4 \beta} \int_{0}^{\beta} d \tau \int d^{D-1} x \Big\{ - \frac{1}{\kappa_{f}} R(\tau,x,z_{f}) + \frac{1}{\kappa_{\varepsilon}} R(\tau,x,\varepsilon) \Big\} . \eqa
Inserting this on-shell free energy of a given disorder realization into
\bqa && \mathcal{F} = \int D g_{\mu\nu}(x,z_{f}) P[g_{\mu\nu}(x,z_{f})] F[g_{\mu\nu}(x,z_{f})] \nonumber \eqa
with the introduction of the power-law distribution function of the metric tensor, we find the disorder-averaged free energy for strongly coupled holographic CFTs.

\section{Discussion}

We believe that our proposal is casting various questions on the physics of disordered strongly interacting conformal field theories in the large central-charge limit. Resorting to the power-law distribution function for the IR boundary metric tensor, one may investigate how the black hole entropy can be quenched-averaged to show a modified area law. Accordingly, one can ask how the quantum chaos from the black hole physics, i.e., the Wigner-Dyson distribution of the level statistics, turns into the physics of Anderson localization or the regime of the Poisson level statistics by this averaging procedure. Moreover, we can discuss how the effective hydrodynamics due to the black hole entropy changes into `rather' integrable quantum dynamics by quenched averaging over correlation functions of metric tensors, gauge fields, and etc.. It is also a fundamental question to ask whether the present theoretical proposal can reproduce the quantum chaos or not.

\begin{acknowledgments}
K.-S. Kim was supported by the Ministry of Education, Science, and Technology (NRF-2021R1A2C1006453 and NRF-2021R1A4A3029839) of the National Research Foundation of Korea (NRF) and by TJ Park Science Fellowship of the POSCO TJ Park Foundation. K.-S. K appreciates fruitful discussions with Shinsei Ryu and Junggi Yoon.
\end{acknowledgments}

%
%
%
%
%
%
%
%

\end{document}